\documentclass[lettersize,journal]{IEEEtran}
\usepackage{amsmath,amsfonts}
\usepackage{algorithmic}
\usepackage{algorithm}
\usepackage{booktabs}       
\usepackage{array}
\usepackage[caption=false,font=normalsize,labelfont=sf,textfont=sf]{subfig}
\usepackage{textcomp}
\usepackage{stfloats}
\usepackage{url}
\usepackage{verbatim}
\usepackage{graphicx}
\usepackage{amsmath,amsfonts,amssymb,amsbsy,fixmath,eucal}
\usepackage{cite}
\usepackage{multirow}
\usepackage{color}

\begin{document}

\title{Attention-based Deep Neural Networks for Battery Discharge Capacity Forecasting}

\author{Yadong Zhang, Chenye Zou and Xin Chen
\thanks{This work was supported in part by the National Natural Science Foundation of China (Grant No. 21773182 (B030103)) and the HPC Platform, Xi'an Jiaotong University. (Corresponding author: Xin Chen, e-mail: xin.chen.nj@xjtu.edu.cn)}
\thanks{Yadong Zhang is with Center of Nanomaterials for Renewable Energy, State Key Laboratory of Electrical Insulation and Power Equipment, School of Electrical Engineering, Xi'an Jiaotong University, Xi'an, Shaanxi, China (e-mail: zhangyadong@stu.xjtu.edu.cn). }
\thanks{Chenye Zou is with Center of Nanomaterials for Renewable Energy, State Key Laboratory of Electrical Insulation and Power Equipment, School of Electrical Engineering, Xi'an Jiaotong University, Xi'an, Shaanxi, China (e-mail: zcy1998@stu.xjtu.edu.cn). }
\thanks{Xin Chen is with Center of Nanomaterials for Renewable Energy, State Key Laboratory of Electrical Insulation and Power Equipment,  School of Electrical Engineering, Xi'an Jiaotong University, Xi'an, Shaanxi, China (e-mail: xin.chen.nj@xjtu.edu.cn). }
}



\maketitle

\begin{abstract}
Battery discharge capacity forecasting is critically essential for the applications of lithium-ion batteries. The capacity degeneration can be treated as the memory of the initial battery state of charge from the data point of view. The streaming sensor data collected by battery management systems (BMS) reflect the usable battery capacity degradation rates under various operational working conditions. The battery capacity in different cycles can be measured with the temporal patterns extracted from the streaming sensor data based on the attention mechanism. The attention-based similarity regarding the first cycle can describe the battery capacity degradation in the following cycles. The deep degradation network (DDN) is developed with the attention mechanism to measure similarity and predict battery capacity. The DDN model can extract the degeneration-related temporal patterns from the streaming sensor data and perform the battery capacity prediction efficiently online in real-time. Based on the MIT-Stanford open-access battery aging dataset, the root-mean-square error of the capacity estimation is 1.3 mAh. The mean absolute percentage error of the proposed DDN model is 0.06{\%}. The DDN model also performance well in the Oxford Battery Degradation Dataset with dynamic load profiles. Therefore, the high accuracy and strong robustness of the proposed algorithm are verified.
\end{abstract}


\section{Introduction}
Forecasting the state of health and lifetime of Li-ion batteries is an unsolved challenge that limits technologies such as consumer electronics and electric vehicles. However, diverse aging mechanisms, significant device variability, and dynamic operating conditions have remained major challenges. Battery Management System (BMS) plays a vital role in integrating many things such as voltage sampling from cell battery, cells balancing, determine State of Charge (SOC), estimate State of Health (SOH), and predict Remaining Useful Life (RUL). Particularly under different operational conditions, the prediction of battery capacity and RUL of lithium-ion batteries is essential in battery health management. The critical technology needed for condition-based maintenance is prognostic and health management. Leading EV battery manufacturers offer customized and smart battery solutions that provide extensive system diagnostics such as accurate cell voltage, state of charge, temperature monitoring, cell balancing, real-time with the help of IoT and data analytics. It enables battery pack manufacturers, OEMs, and electric mobility fleet operators to leverage smart edge-computing hardware and data-driven AI models to obtain health insights, constantly monitor and improve the life and performance of batteries.

The new engineering approach that allows a real-time assessment of the system's health online becomes possible with the development of the data-driven model. The cycle life of batteries is the number of charge and discharge cycles that a battery can complete before losing performance. The data for real-time sensor stream of the time-dependent characteristic information of the battery demand the development of a more efficient and reliable algorithm to forecast capacity degeneration. The data-driven machine learning models have been developed for the SOC, SOH and the capacity degeneration prediction of lithium-ion batteries\cite{hannan2017review,lipu2018review}. The battery degeneration and remaining useful life are strongly path-dependent. The capacity degeneration is reflected in the behavior change of working load streaming sensor data of the battery. Therefore, the historical behaviors in the streaming sensor data collected by the field sensors are strongly important to capacity degeneration forecasting.

Given the path dependence in capacity degeneration, RNN (including LSTM) is important in many previous works. In 2012, the research found that RNN can use to monitor the SOH\cite{eddahech2012behavior}. They chose several features such as temperature, current pulse magnitude, and SOC variations. Later on, LSTM combined with Monte Carlo simulation was used to fit capacity degeneration curve and get probabilistic RUL prediction in 2018\cite{zhang2018long}. The next year, an Elman-LSTM model was proposed to predict RUL\cite{li2019remaining}. Elman model worked as a high-frequency sub-layer, and LSTM worked as a low-frequency sub-layer. Apart from RNN and LSTM, supporting vector machine (SVM) is also widely used. Some researchers took sample entropy as a feature and used SVM and RVM to predict SOH\cite{widodo2011intelligent}. In 2017, research showed that supporting vector regression (SVR) and particle filter (PF) could be combined with each other to fit the capacity fade curve and predict the RUL\cite{wei2017remaining}. Different features, such as the time interval of an equal charge voltage difference (TIECVD) and the time interval of an equal discharge voltage difference (TIEDVD), could also be fed into the SVR model \cite{zhao2018novel}. Gaussian process regression can be used to predict battery calendar aging instead of cycling aging\cite{liu2019gaussian}. Random forest can predict battery lifetime in IoT devices\cite{maddikunta2020predictive}. Some researches also showed that Deep Neural Networks (DNN)\cite{khumprom2019data} had a better performance than kNN, LR, SVM, and ANN while predicting SOH and RUL.

Recently, many works have been focusing on capacity degeneration forecasting problems. Mainly, machine learning methods attract a lot of attention. The SBC-RBFNN-mRFR-based scheme is proposed for offline and online SOH estimation. It has good accuracy for the laboratory dataset and performs well for the real-world EV dataset\cite{she2021offline}. Based on the comprehensive battery degradation dataset with different cycling conditions, a nonlinear autoregressive exogenous (NARX) model is proposed to get the relationship between SOH and the features extracted from the partial charging voltage curve\cite{khaleghi2021online}. Temporal convolutional network (TCN) is also used in capacity forecasting. The empirical mode decomposition (EMD) technology is applied to denoise the data to reduce the impact of local regeneration, and then TCN is used to estimate SOH and RUL\cite{zhou2020state}. Furthermore, three feature selection methods and four machine learning methods are combined to test SOH prediction performances. Then they found that the fusion-based selection method and Gaussian process regression (GPR) had an overall better result than other combinations\cite{hu2020battery}.

The key to forecasting and improving battery life lies in extracting useful feature information from the streaming sensor data collected by the field sensors of BMS. BMS monitors the battery with the streaming sensor data of various items, including voltage, current, temperature, and coolant flow, collected with the field sensors. Therefore, data analytics are essential to model the battery SOC, SOH, state of power (SOP), and {\it{etc.}}. To directly forecast the battery capacity degeneration with streaming sensor data demands the development of a new algorithm. The temporal patterns in the streaming data are strongly associated with battery degeneration. Streaming sensor data is very similar to the sequence of words in the text. Attention mechanism originates from Neural Machine Translation (NMT) field\cite{NMT}. NMT takes a weighted sum of all the annotations to get an expected annotation and focuses only on information relevant to the generation of the next target word. The attention mechanism doesn't need prior knowledge for training sets of representative data. Feature similarity is useful for high-level object category detection and classification in computer vision. The attention-based model can capture relative interests with feature similarity. Choosing informative, discriminating and independent features is crucial. Feature-based attention prioritizes the processing of non-spatial features across the visual field. Recent work has applied the attention mechanism to the click-through rate prediction in online advertising \cite{ali} by adaptively learning the representation of user interests from historical behaviors. We develop the Deep Degradation Network (DDN) to forecast capacity degeneration. 

The DDN model is proposed to extract the temporal feature from the streaming data based on the attention mechanism, catch the capacity degeneration, and make the capacity prediction. The streaming sensor data at each cycle is weighted according to the attention-based distance from the initial cycle. The DDN model can catch the temporal features in battery capacity degeneration.

The paper is organized into four sections. In Section~\ref{DDN}, we discuss the meaning of the capacity degeneration in terms of the attention mechanism. The DDN model is presented. In Section~\ref{data}, we discuss the data sources and the preprocessing of the data corresponding to the features. In Section~\ref{deep}, we discuss the experiments and results. The concluding remarks are given in Section~\ref{conclusion}.

\section{Theory and Model: Deep Degradation Network}\label{DDN}

\begin{figure}[H]
  \centering
  \includegraphics[width=0.5\textwidth]{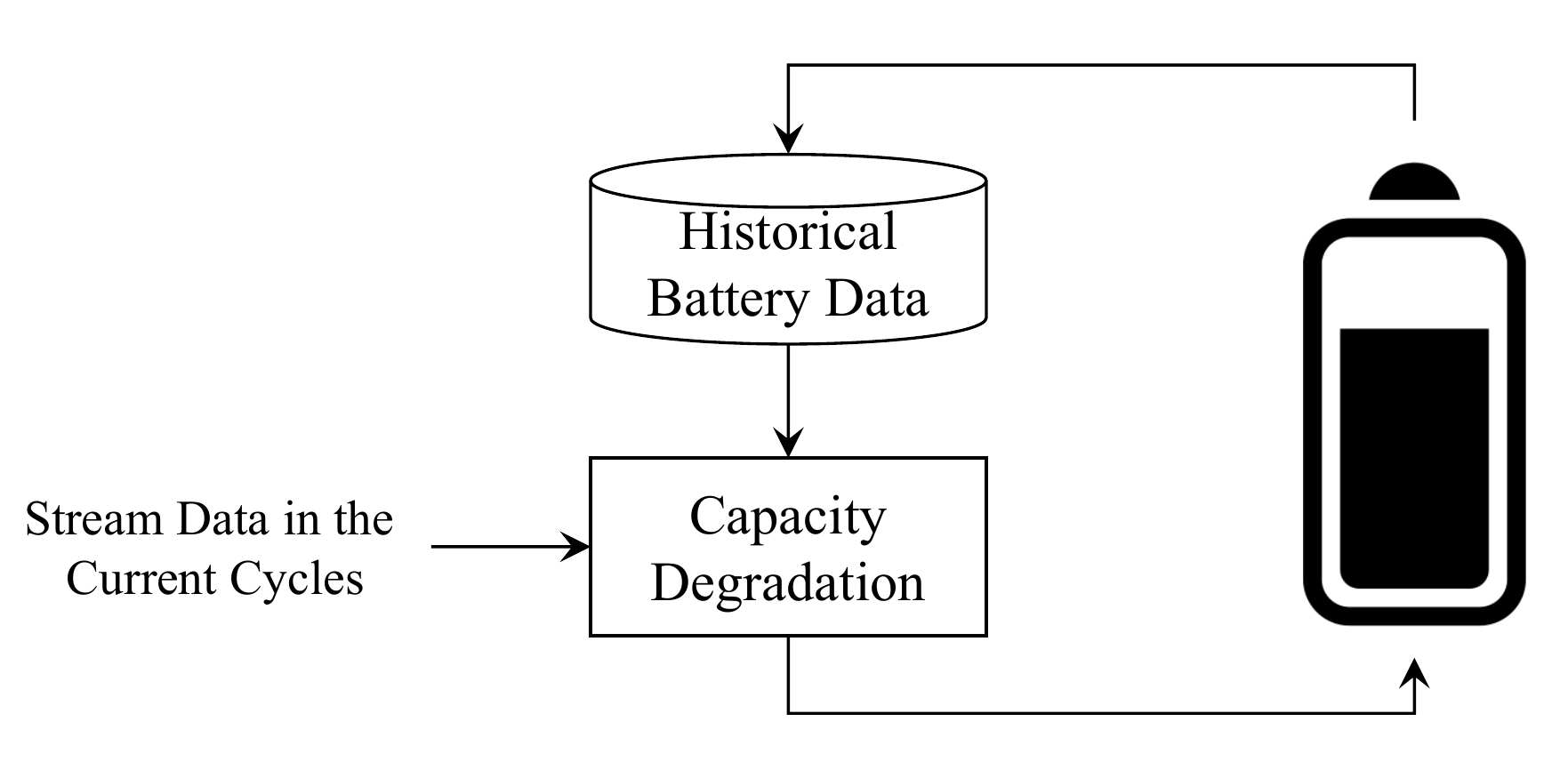}
  \caption{Illustration for the battery capacity forecasting based on the similarity of battery performance between the current and initial cycles with the attention mechanism}
  \label{fig:similar}
\end{figure}

The streaming sensor data contains all the dynamic information about the SOC, SOH, and battery capacity. The similarity of battery performance between the current and initial cycles can characterize capacity degeneration. Given the features about the battery capacity are embedded in the streaming sensor data, Fig.~\ref{fig:similar} shows that the degeneration can be measured as the matching of the features embedded in the streaming data in the battery cycles with the initial cycle. The attention mechanism emerges naturally from problems that deal with time-series data. Dealing with “sequences” in time-series demands a new formulation of the problem in machine learning first. Inspired by the deep internet network (DIN) model\cite{ali}, we propose the DDN by inheriting the attention unit used in the DIN model. In the DDN model, the attention mechanism measures the battery capacity in the battery cycles. 

\subsection{Feature-based Attention and Capacity Degeneration}
Similarity measures are central to pattern recognition in the streaming sensor data. Dynamic pattern recognition can be used to describe the evolution of complex nonlinear dynamics. In processing a high volume of streaming sensor data, informative features count a great deal in the attention mechanism. Attention became popular in the general task of coping with the streaming sensor data. The attention mechanism was developed originally to help memorize long sentences in NMT, which is the measurement of semantic similarity. LiFePO4 lithium-ion batteries still have path-dependent memory effects\cite{sasaki2013memory}. The nonlinear and complex dynamic memory effect is reflected in the streaming sensor data collected by the BMS. As a result, the battery capacity degeneration is path-dependent in the charge and discharge cycles, given the battery operational working conditions.

\begin{figure*}[ht]
  \centering
  \subfloat[]{\includegraphics[width=0.40\textwidth]{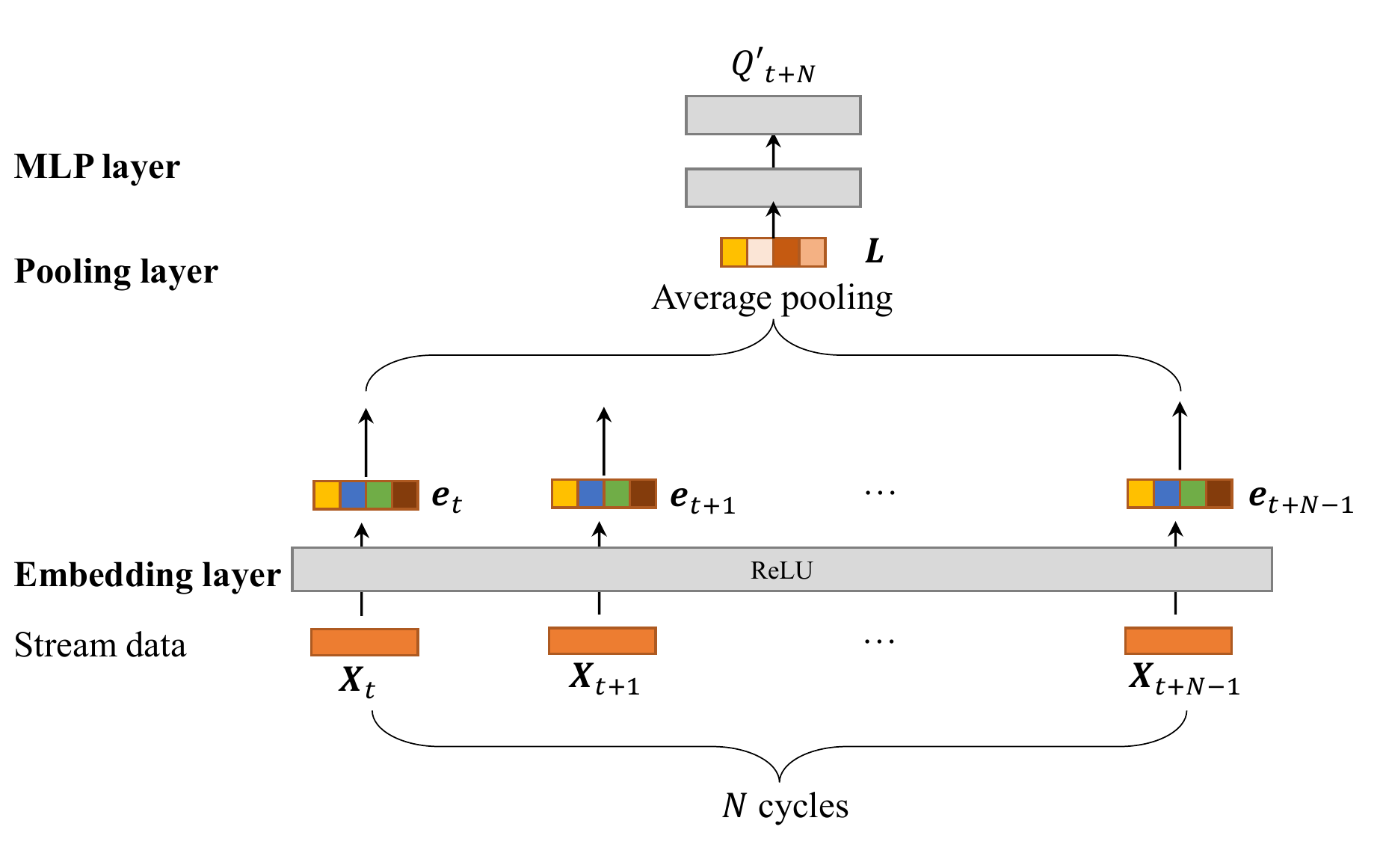}\label{fig:basemodel}}
  \subfloat[]{\includegraphics[width=0.60\textwidth]{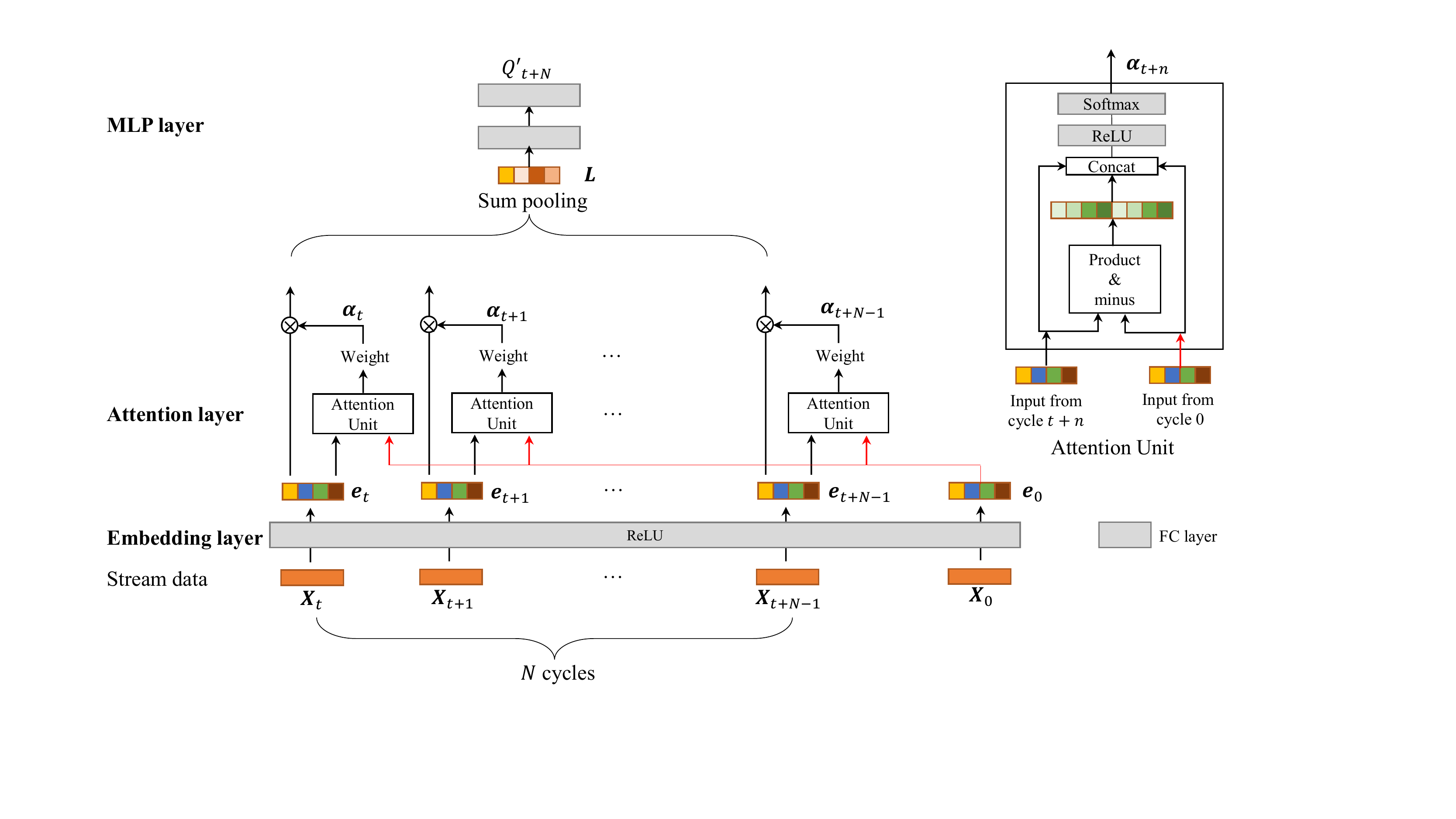}\label{fig:framework}}
  \caption{The scheme of the DDN model for the battery capacity prediction} \label{fig:model}
\end{figure*}

The battery capacity degeneration, {\it{i.t.}} the level of capacity for the battery to hold charges, is reflected in the pattern change of operational streaming sensor data and can be measured by the similarity between the current cycle and the initial cycle. The attention mechanism prioritizes the processing of sensory information at specific spatial locations  or with specific feature values. Therefore, the attention mechanism can predict battery capacity degradation with the attention-based similarity measurement based on the selected features from the streaming sensor data in the different cycles. DIN is a state-of-the-art model to use attention mechanism to capture user interests from historical behaviors\cite{ali}. From the data perspective, the battery operational streaming sensor data and user historical behavior data are very similar. Therefore, the attention mechanism is embedded into the sequential structure. The battery historical operational data is critically important to predicting the battery capacity in each cycle since the states of lithium-ion batteries have the historical path dependence. With the selected features of the battery operational streaming sensor data, similarly, the DDN model is proposed to use an attention mechanism to capture the pattern change of the operational streaming sensor data acquired by the sensors in the BMS for the battery capacity degradation predictions. The DDN model takes the basic structure of the deep interest network. By extracting the degradation patterns from battery operational data, including the voltage, current, impedance, and {\it{etc.}}, DDN can model the temporal interest and attention-based similarity in different cycles. The DDN model consists of three layers, embedding layer, attention layer, and MLP output layers in Fig.~\ref{fig:framework}. The attention mechanism is realized in the attention unit. For a battery, the similarity between the current cycle and the initial cycle can be measured based on the streaming sensor data.

\subsection{Structure of Deep Degradation Network}

{\it \textbf{Embedding layer.}} The base structure of DDN is shown in Fig.~\ref{fig:basemodel}. The embedding layer is used to transform the streaming sensor data in the $N$th cycles from $\mathbold{X}_t$ to $\mathbold{X}_{t+N-1}$ and the initial cycle $\mathbold{X}_0$ into the low dimensional dense representation,
\begin{equation}
  \mathbold{E}_t=[\mathbold{e}_{t}, \mathbold{e}_{t+1}, \cdots, \mathbold{e}_{t+n}, \cdots, \mathbold{e}_{t+N-1}] \in \mathbb{R}^{D \times N}
\end{equation}
where $\mathbold{e}_{t+n}$ is the $D$-dimension embedding encoded from the streaming sensor data in the $t+n$th cycle.

{\it \textbf{Pooling layer.}} By transforming the historical embedding via a mean pooling layer to get the dense representation $\mathbold{L}$ for MLP:
\begin{equation}
  \mathbold{L}= \frac{1}{N}\sum_{n=0}^{N-1}{\mathbold{e}_{t+n}} \in \mathbb{R}^{D}
\end{equation}

{\it \textbf{MLP.}} Given the dense representation vector $\mathbold{L}$, the predicted capacity $\hat{Q}$ at the $(t+N)$th cycle are obtained with a MLP including the hidden and output fully connected layers,
\begin{align}
  \mathbold{o} & = \mathbold{W}_o \mathbold{L} + \mathbold{b}_o,                  \\
  \hat{Q}_{t+N}     & =  \mathbold{W}_q \mathbold{o}  + \mathbold{b}_q \in \mathbb{R},
\end{align}
where $\mathbold{W}_o$ and  $\mathbold{W}_q$ are the weights matrix for the hidden and output layers and $\mathbold{b}_o$ and $\mathbold{b}_q$  bias.

 {\it \textbf{Loss.}} The loss function is the mean-squared-error function,
\begin{equation}
  Loss = \frac{1}{M}\sum_{i=1}^B\sum_{t=0}^{T_i} (\hat{Q}_{t+N} - Q_{t+N})^2
\end{equation}
where B is the number of batteries in the training set, $T_i$ is the number of capacities in cycles for the $i$th battery,  $M=\sum_{i=1}^B T_i$, $\hat{Q}$ the predicted capacity, and $Q$ the measured discharge capacity.

\subsection{Similarity Attention and Capacity Degeneration}
Attention is strongly associated with memory and development. From the streaming sensor data point of view, the battery degeneration can be treated as the memory of the initial battery state of charge. In the base structure, the traditional global average pooling of the feature embeddings is used. To evaluate the development of battery degradation, we propose degradation attention to assess the degradation weights of feature embeddings in reference to the embedding of the initial cycle. The attention unit can measure the degradation memory in the current cycle as to the initial cycle in terms of feature embedding extracted from the streaming sensor data. The degradation weights of the embeddings as the output of the attention unit define the attention-based similarity with regards to the embedding of the initial cycle. As a result, the pooling layer with the attention weight is,
\begin{equation}
  \mathbold{L}= \sum_{n=0}^{N-1}{\alpha_{t+n} \mathbold{e}_{t+n}} \in \mathbb{R}^{D}
\end{equation}
where $\alpha_{t+n}$ is the attention weight from the output of the attention unit.

The embedding of the initial cycle are the reference for the measurement of the degradation development in the cycles. This makes the network learn the degradation patterns. In order to measure the difference of embeddings between the current and the initial cycles, the concatenation of the embeddings is defined as,
\begin{align}
  \mathbold{w}_{t+n} = [\mathbold{e}_{t+n}; \mathbold{e}_{0}; \mathbold{e}_{t+n} - \mathbold{e}_{0}; \mathbold{e}_{t+n} \mathbold{e}_{0}]
\end{align}
The attention unit in Fig.~\ref{fig:model} has the two fully connected hidden layers with the rectifier linear unit (ReLU) and softmax.
\begin{align}
  \mathbold{h}_{t+n} & = ReLU(\mathbold{W}_h \mathbold{w}_{t+n}  + \mathbold{b}_h) \\
  {z}_{t+n}          & = \mathbold{W}_z \mathbold{h}_{t+n} + \mathbold{b}_z
\end{align}
The final output of the attention weight is defined as the softmax of ${z}_{t}$ as a normalization of ${z}_{t+n}$ over the historical $N$ cycles starting at $t$,
\begin{align}
  \alpha_{t+n} & = softmax(z_{t+n}) = \frac{e^{z_{t+n}}}{\sum_{k=0}^{N-1}{e^{z_{t+k}}}}
\end{align}

\subsection{Battery Capacity Features}
The streaming sensor time-series data contain rich information about capacity degradation. The four typical features, including voltage, current, historical discharged capacity, and impedance, are described in Table~\ref{tab:datatype}. The four features are labeled as $1,2,3,4$ for the voltage, current, historical discharged capacity, and impedance accordingly in the DNN model. 
\begin{table}[H]
  \caption{Feature sets of the streaming sensor data measured by BMS.}
  \centering
  \begin{tabular}{lll}
    \toprule
    Operational data                  & Type    & Unit          \\
    \midrule
    1. Voltage                        & float   & V             \\
    2. Current                        & float   & A             \\
    3. Historical Discharged Capacity & float   & Ah            \\
    4. Impedance                      & complex & $\rm{\Omega}$ \\
    \bottomrule
  \end{tabular}
  \label{tab:datatype}
\end{table}

For the DDN model, the streaming sensor data at the $(t+n)$th cycle, $\mathbold{X}_{t+n}$ needs to be encoded into the embedding for each feature $j, j=\{1,2,\cdots, J\}$. The streaming sensor data of the $j$th feature at the $(t+n)$th cycle, $\mathbold{X}_{t+n}^{(j)}$ is encoded into $\mathbold{e}_{t+n}^{(j)}$ with the fully connected layer,
\begin{equation}
  \mathbold{e}_{t+n}^{(j)} = \mathbold{W}_{e}^{(j)}\mathbold{X}_{t+n}^{(j)} + \mathbold{b}_{e}^{(j)}
\end{equation}
where $\mathbold{W}_{e}^{(j)}, \mathbold{b}_{e}^{(j)}$ are the weight and bias of the $j$th fully connected layer. The length of streaming sensor data $\mathbold{X}_{t+n}^{(j)}$ is $l^{(j)}$.
The dimensions of encoded embeddings of the $j$th category is $K^{(j)}$. The concatenated encoded embeddings of all the $J$ categories for the $t+n$ cycle is,
\begin{equation}
  \mathbold{e}_{t+n} = [\mathbold{e}_{t+n}^{(1)}; \mathbold{e}_{t+n}^{(2)};  \cdots; \mathbold{e}_{t+n}^{(J)}]
\end{equation}
where $J$ is the number of features and $D = \sum_1^J K^{j}$ the total size of $\mathbold{e}_{t+n}$. 
For the DDN model, the streaming sensor data in the first cycle $\mathbold{X}_0$ is used as the reference for the capacity degradation prediction. The streaming sensor data in the following cycles are used to train the DDN model. Since the DDN model use the $N$ historical cycles, so the sequence of the streaming sensor data $\mathbold{X}_t$ are converted into a sequence of moving frames $[\mathbold{X}_t, \cdots, \mathbold{X}_{t+N-1}]$.

The charge and discharge voltage curves are the critical information to the capacity prediction in the charge and discharge processes. The terminal voltage of the lithium-ion battery can be represented as:
\begin{align}
  V_C =E_{+}-E_{-}+V_R ,\\
  V_D =E_{+}-E_{-}-V_R ,
\end{align}
where, $V_C$ and $V_D$ are the terminal voltages for the charge and discharge cycle respectively. $E_{+}$ denotes the anode potential, $E_{-}$ the cathode potential, and $V_R$  the voltage difference caused by the polarization. According to the Nernst equation, $E_{+}$ and $E_{-}$ are related to the electrode material, temperature, and concentration of lithium-ions. In addition, the internal Ohmic resistances in batteries influence $V_R$. As the battery capacity degenerates, the charge voltage curve will rise faster, and discharge voltage curve will drop faster because the battery internal resistance increases. As a result, the charge and discharge voltage curves can be used as features to predict battery degradation.

\section{Data Sources}\label{data}
Three typical battery datasets are used for the battery capacity predictions. The first one is the NASA PCoE dataset\cite{data1,goebel2008prognostics} with 38 lithium-ion batteries. For the NASA PCoE dataset, $\mathbold{X}_t$ contains the discharge capacity, charge, and discharge voltage data. We take the first 1500 seconds of the charge and discharge voltage curves, respectively, in each cycle. Since the actual sampling rate is not constant for the charge and discharge voltage curves, we need to unify the sizes of the two voltage curves for the DDN embedding layer. The new curves are linearly interpolated and re-sampled with $l^{(2)}$ and $l^{(3)}$ uniform sampling points for the charge and discharge curves, respectively.

In the following experiments, we take two tests for the NASA PCoE datasets, the PCoE$^1$ and PCoE$^2$ datasets used in the benchmark methods of the SVR-based model\cite{wei2017remaining} and the sampEN model\cite{li2019remaining}. For the NASA PCoE$^1$ dataset, these batteries run through 3 different operational profiles (charge, discharge, and impedance) at room temperature. They were charged with 1.5A CC and 4.2V CV and discharged with 2A CC. Impedance measurement was carried out through electrochemical impedance spectroscopy (EIS) frequency sweep from 0.1Hz to 5kHz. For the NASA PCoE$^2$ dataset, these batteries run through 3 different operational profiles (charge, discharge, and impedance) at different temperatures and discharging profiles. Charging and impedance profiles are the same as the NASA PCoE$^1$ dataset.

The second one is the the MIT-Stanford open-access dataset \cite{severson} with 124 commercial lithium-ion batteries under the fast-charging conditions. For the MIT-Stanford open-access dataset, all the batteries manufactured by A123 Systems (APR18650M1A) were cycled in horizontal cylindrical fixtures on a 48-channel Arbin LBT potentiostat in a forced convection temperature chamber set at $30^\circ$C. These batteries have different charging profiles so that they age at different rates. Each cell has the 1.1Ah nominal capacity and the 3.3V nominal voltage.

For the MIT-Stanford open-access dataset, $\mathbold{X}_t$ contains the streaming sensor data of the discharge capacity, charge voltage, and discharge voltage for each cycle. The length of historical capacity data is $l^{(1)}$. We take the first 6 minutes of the charge and discharge voltage curves respectively in each cycle. Also, they are linearly interpolated and re-sampled with $l^{(2)}$ and $l^{3)}$ uniform sampling points for the charge and discharge voltage curves.

We divide the data into three datasets, the training dataset (75\%), the validation dataset (10\%), and the test dataset (15\%). First, the indices of all the batteries are shuffled. Then, the first 75\% of the shuffled indices are training set. The following 10\% of the shuffled indices are validation set, and the remaining 15\% of the shuffled indices are testing set. The DDN model predicts the capacity in the $t+N$th cycle with the streaming sensor data from the  $t$th to $t+N-1$th historical cycles. The streaming sensor data in the N historical cycles from $\mathbold{X}_t$ to $\mathbold{X}_{t+N-1}$ is the moving frames to feed into the embedding layer in the DDN model. The number of cycles measured for each battery and the number of historical cycles $N$ decide the number of moving frames.

Batteries in the above two datasets were cycled in the ideal condition. They were discharged at a constant C-rate and were always fully charged and discharged at each cycle. To investigate the performance of the DDN model under dynamic load profiles, we use the Oxford Battery Degradation Dataset\cite{birkl2017oxford, birkl2017diagnosis}. This dataset contains eight lithium-ion batteries, which are tested under $40^\circ$C. These batteries were charged with a CC-CV charging profile and discharged with the urban Artemis driving profile. 
After every 100 drive cycles, characterization tests were performed, which included a 1C cycle and a C/18 pseudo-OCV cycle, to get the information of the ideal charge and discharge profile and the OCV curve. The dataset only contains characterization cycles, and drive cycles are not recorded. 
To benchmark the performance with the SBC-RBFNN model\cite{she2021offline}, we divide the batteries in the dataset into four cases as used for the training of the SBC-RBFNN model. The specification of how to divide the dataset into the training and test datasets are given in Section~\ref{deep}.


\section{Experiments}\label{deep}
\label{sec:others}
Two performance tests in ideal condition are carried out with the NASA PCoE and MIT-Stanford open-access datasets. In the first performance test, we test the performance of the DDN model with the small-sized NASA PCoE$^1$ and PCoE$^2$ dataset in benchmark with the SVR-based method\cite{wei2017remaining} and the sample-entropy method\cite{li2019remaining}. The two benchmarks with the SVR-based and sample-entropy use the different NASA PCoE datasets. In the second performance test, we test the performance of the DDN model with the large-sized MIT-Stanford open-access dataset. The performance of the capacity prediction in each cycle is evaluated by the area under the Root Mean Squared Error (RMSE), Mean Absolute Percentage Error (MAPE), and R-squared (R$^2$) according to the moving historical cycles on the test dataset.
\begin{align}
  RMSE = \sqrt {\frac{1}{M}\sum_{i=1}^B\sum_{t=0}^{T_i} (\hat{Q}_{t+N} - Q_{t+N})^2} \\
  MAPE = \frac{1}{M}\sum_{i=1}^B\sum_{t=0}^{T_i} \frac{|\hat{Q}_{t+N} - Q_{t+N}|}{Q_{t+N}} \times 100\%
\end{align}
where $B$ is the number of batteries in the test dataset, $T_i$ is the number of predicted capacities in cycles for the $i$th battery, $M=\sum_{i=1}^B T_i$, $\hat{Q}$ the predicted capacity, and $Q$ the discharge capacity.

In the first performance test, we use No. 6 battery for the training of the DDN model, and  No. 5, No. 7, and No. 18 three batteries for the test as used in the SVR-based method\cite{wei2017remaining}.
The streaming sensor data of the charge and discharge voltages are normalized in the following way:
\begin{align}
  V_{discharge} = \frac{4.2 - V_{discharge}}{4.2} \\
  V_{charge} = \frac{V_{charge}}{4.2}
\end{align}
where $V_{discharge}$ is the discharge voltage data, $V_{charge}$ is the charge voltage data. Noting that for the historical capacity data, we take the same normalization way as the capacity time-series data which is illustrated above.
At the same time, we normalize the capacity time-series data with a maximum value of 2.1 and a minimum of 1.1 in the moving frames. The hyper-parameters of the DDN model are given in Table~\ref{tab:params}. 
Then, with mean squared error as loss function, the network is trained with Adam at the learning rate 0.001, $\beta_1$ 0.9 and $\beta_2$ 0.999. The performances of the DDN model and SVR-based method is shown in Table~\ref{tab:result-nasa-rmse}. 

On the other hand,  we also use No. 34, No. 36, and No. 51 three batteries for the training of the DDN model, and No. 27, No. 31, and No. 55 three batteries for the test as used the sample-entropy methods\cite{li2019remaining}. The six batteries have the streaming sensor data measured at the different time-varying temperatures. Therefore, we do not normalize the capacity time-series data like the first performance test. 
The hyper-parameters of the DDN model are presented in Table~\ref{tab:params}. 
With the mean squared error as the loss function, the network is trained with Adam at the learning rate 0.001, $\beta_1$ 0.9 and $\beta_2$ 0.999. The comparison of the DDN and SampEN models is shown in Table~\ref{tab:result-nasa-mape}.

\begin{table}[H]
  \caption{The performance of DDN compared to the SVR-based Model\cite{wei2017remaining}}
  \centering
  \begin{tabular}{lllll}
    \toprule
    Name                        & 5      & 7      & 18     \\
    \midrule
    DDN  (RMSE (Ah))            & 0.0147 & 0.0132 & 0.0237 \\
    SVR-based model (RMSE (Ah)) & 0.0146 & 0.0147 & 0.0229 \\
    \bottomrule
  \end{tabular}
  \label{tab:result-nasa-rmse}
\end{table} 
\begin{table}[H]
  \caption{The performance of the DDN Model compared to the SampEN Model\cite{li2019remaining}}
  \centering
  \begin{tabular}{lllll}
    \toprule
    Name          & 31     & 55     & 27     \\
    \midrule
    DDN  (MAPE)   & 0.92\% & 2.71\% & 0.93\% \\
    SampEN (MAPE) & 0.74\% & 1.7\%  & 1.38\% \\
    \bottomrule
  \end{tabular}
  \label{tab:result-nasa-mape}
\end{table}

In the second performance test, we use the large-sized MIT-Stanford open-access dataset to train the DDN model. We normalize the capacity time series data with a maximum value of 1.1 and a minimum value of 0.8. Then, with mean squared error as loss function, the network is trained with Adam at the learning rate 0.001, $\beta_1$ 0.9 and $\beta_2$ 0.999, and it is validated with the early stopping criteria on the validation set. The streaming sensor data of the charge and discharge voltages are normalized as follows,
\begin{align}
  V_{discharge} = \frac{3.2 - V_{discharge}}{3.2} \\
  V_{charge} = \frac{3.6 - V_{charge}}{3.6}
\end{align}
where $V_{discharge}$ is the discharge voltage data, $V_{charge}$ is the charge voltage data. Noting that for the historical capacity data, we also normalize them with a maximum value of 1.1 and a minimum value of 0.8. The hyper-parameters of the DDN model are listed in Table~\ref{tab:params}. 

\begin{figure}[H]
  \subfloat[The capacity degeneration curves for the 19 batteries in the test dataset in the MIT-Stanford open-access dataset]{\includegraphics[width=0.48\textwidth]{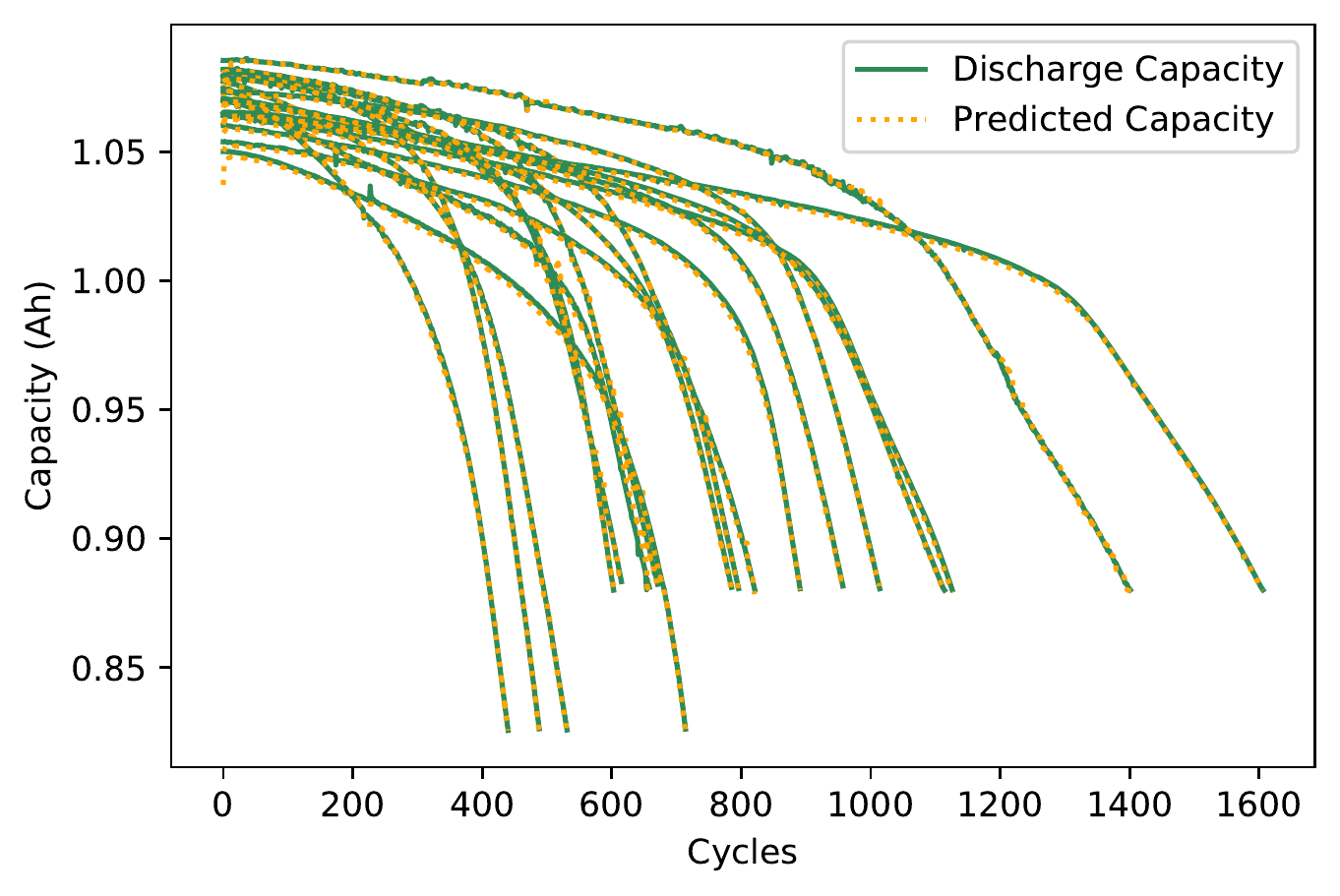}\label{fig:batteries}}\\
  \subfloat[MAPE for 19 batteries in cycles]{\includegraphics[width=0.48\textwidth]{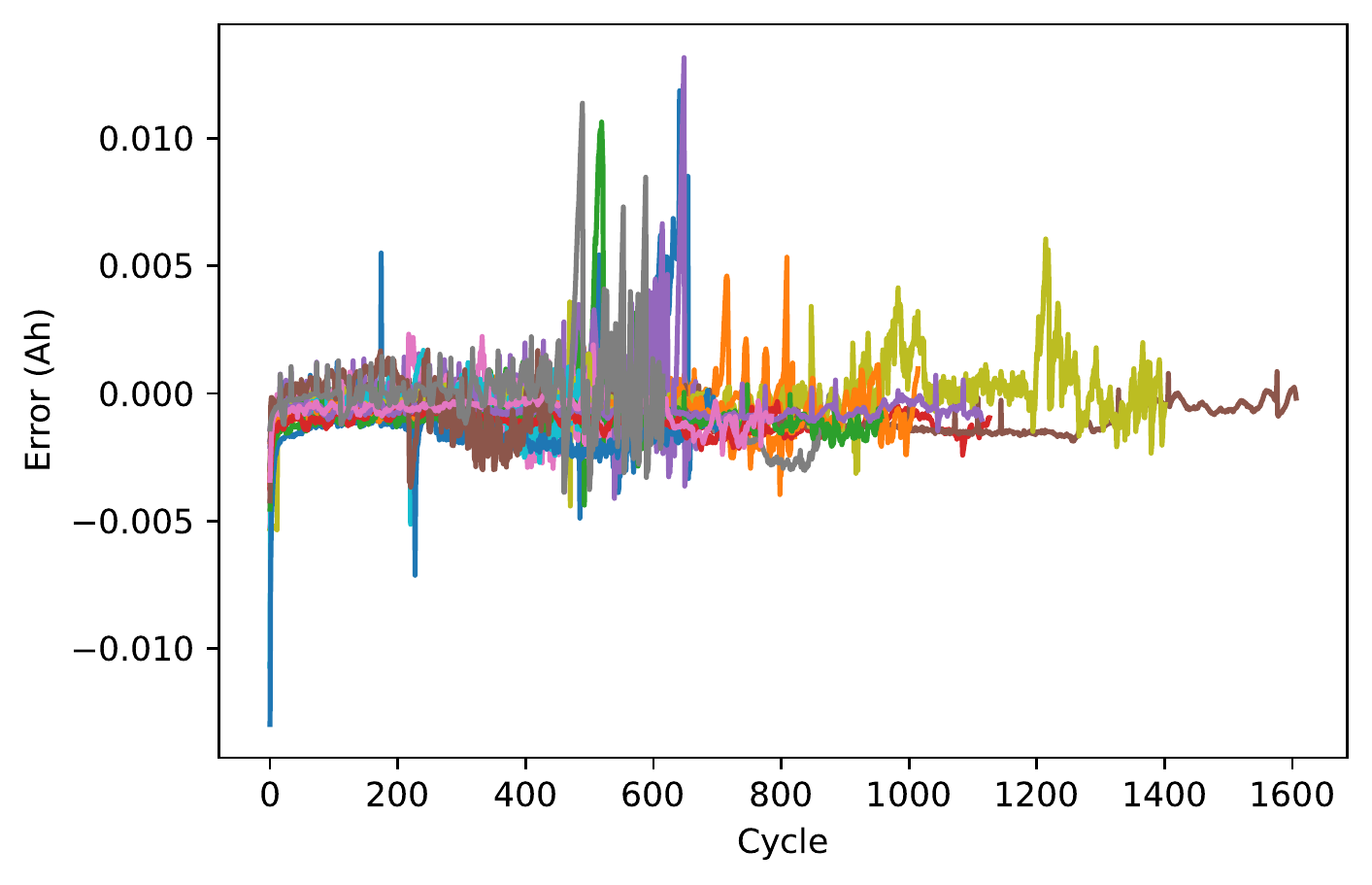}\label{fig:cycles}}
  \caption{The capacity degeneration curves for the 19 batteries in the test dataset in the MIT-Stanford open-access Dataset. The green lines are the real capacity degeneration curves and red dots are the predicted capacity degeneration curves with the DDN model}\label{fig:method}
\end{figure}

Fig.~\ref{fig:method} show the comparison of the discharge capacity degeneration curves and the predicted capacity degeneration curves for the 19 batteries in the test dataset in MIT-Stanford open-access battery aging dataset. The performance of the DDN model is shown in Table~\ref{tab:result-fed}. Based on the MIT-Stanford open-access battery aging dataset, the root-mean-square error (RMSE) of he capacity estimation is 1.3 mAh. The mean absolute percentage error (MAPE) of the predicted capacities for the 19 batteries in the test dataset is 0.06{\%}. 
\begin{table}[H]
  \caption{The performance of the DDN model}
  \centering
  \begin{tabular}{llll}
    \toprule
    Name & RMSE (Ah) & MAPE (\%) & R$^2$  \\
    \midrule
    DDN  & 0.0013    & 0.0626    & 0.9993 \\
    \bottomrule
  \end{tabular}
  \label{tab:result-fed}
\end{table}

\begin{figure}[H]
  \centering
  \includegraphics[width=0.45\textwidth]{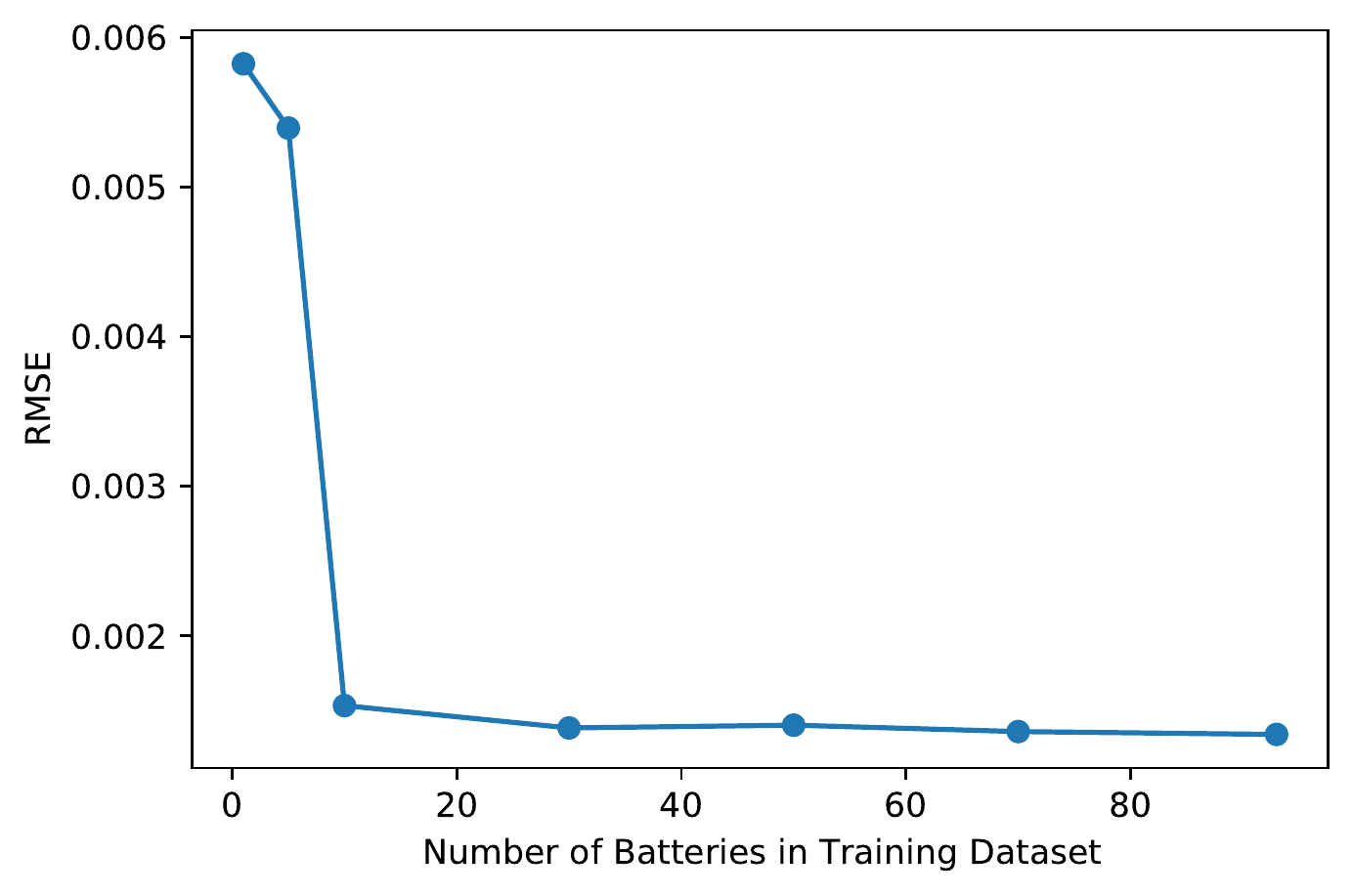}
  \caption{The performance of the DDN model with the training datasets of different sizes in terms of RMSE.}\label{fig:FL_compare}
\end{figure}

We also examine how the performance of the DDN model depends on the size of the training dataset. Fig.~\ref{fig:FL_compare} shows that the DDN model trained based on the 10 battery dataset already demonstrates the same performance as the 93 battery training dataset.

\begin{table*}[ht]
  \caption{Hyper-parameters for the experiments}
  \centering
  \begin{tabular}{llllll}
    \toprule
    \multirow{2}*{Notation}  & \multicolumn{4}{c}{Dataset} & \multirow{2}*{Note}\\
    ~ & NASA PCOE$^1$ & NASA PCOE$^2$ & MIT-Stanford & Oxford & ~\\
    \midrule
    $N$       & 3 & 3 & 30 & 3 & Number of historical cycles                             \\
    $l^{(1)}$ & 1 & 1 & 1 & 1 & Length of historical capacity                           \\
    $l^{(2)}$ & 300 & 300 & 300 & 300 & Length of charge voltage curve                        \\
    $l^{(3)}$ & 300 & 300 & 300 & 300 & Length of discharge voltage curve                     \\
    $K^{(1)}$ & 64 & 64 & 64 & 64 & Dimension of embeddings for historical capacity         \\
    $K^{(2)}$ & 64 & 64 & 64 & 64 & Dimension of embeddings for charge voltage curve      \\ 
    $K^{(3)}$ & 64 & 64 & 64 & 64 & Dimension of embeddings for discharge voltage curve   \\
    $H_1$     & 64 & 64 & 64 & 64 & Dimension of $\mathbold{o}$ in the MLP                  \\
    $H_2$     & 128 & 128 & 128 & 128 & Dimension of $\mathbold{h}_{t+n}$ in the attention unit \\
    Min       & 1.1 & N/A$^*$ & 0.8 & 0.75 & the minimum value for the min-max-normalization         \\
    Max       & 2.1 & N/A$^*$ & 1.1 & 1 & the maximum value for the min-max-normalization         \\
    \bottomrule
  \end{tabular}\\
  $*$ there is no min-max-normalization implemented for the time series in the moving frames.
  \label{tab:params}
\end{table*}

\begin{table*}[ht]
  \caption{The RMSE performance of the DDN model compared to the SBC-RBFNN model\cite{she2021offline}}
  \centering
  \begin{tabular}{lllllll}
    \toprule
    Training Battery Set & \multicolumn{5}{c}{RMSE of Test Battery} & Average RMSE  \\
    \midrule
    Case 1 (No. 2, 3, 8) & No. 1 & No. 4 & No. 5 & No. 6 & No. 7 & ~\\
    DDN & 0.00222 & 0.00371 & 0.00223 & 0.00398 & 0.00204 & 0.00284\\
    SBC-RBFNN & 0.00716 & 0.00421 & 0.00293 & 0.00374 & 0.00727 & 0.00506\\
    \hline
    Case 2 (No. 3, 4, 6) & No. 1 & No. 2 & No. 5 & No. 7 & No. 8 & ~\\
    DDN & 0.00200 & 0.00212 & 0.00210 & 0.00322 & 0.00240 & 0.00237\\
    SBC-RBFNN & 0.00613 & 0.00785 & 0.00322 & 0.00889 & 0.00697 & 0.00661\\
    \hline
    Case 3 (No. 1, 2, 7) & No. 3 & No. 4 & No. 5 & No. 6 & No. 8 & ~\\
    DDN & 0.00223 & 0.00404 & 0.00224 & 0.00421 & 0.00202 & 0.00295\\
    SBC-RBFNN & 0.00559 & 0.00471 & 0.00305 & 0.00368 & 0.00762 & 0.00493\\
    \hline
    Case 4 (No. 1, 7, 8) & No. 2 & No. 3 & No. 4 & No. 5 & No. 6 & ~\\
    DDN & 0.00232 & 0.00207 & 0.00377 & 0.00220 & 0.00408 & 0.00289\\
    SBC-RBFNN & 0.00797 & 0.00470 & 0.00422 & 0.00330 & 0.00393 & 0.00482\\
    \bottomrule
  \end{tabular}
  \label{tab:result-oxford}
\end{table*}

\begin{figure*}[ht]
  \centering
  \subfloat[The capacity degeneration for Case 1]{\includegraphics[width=0.45\textwidth]{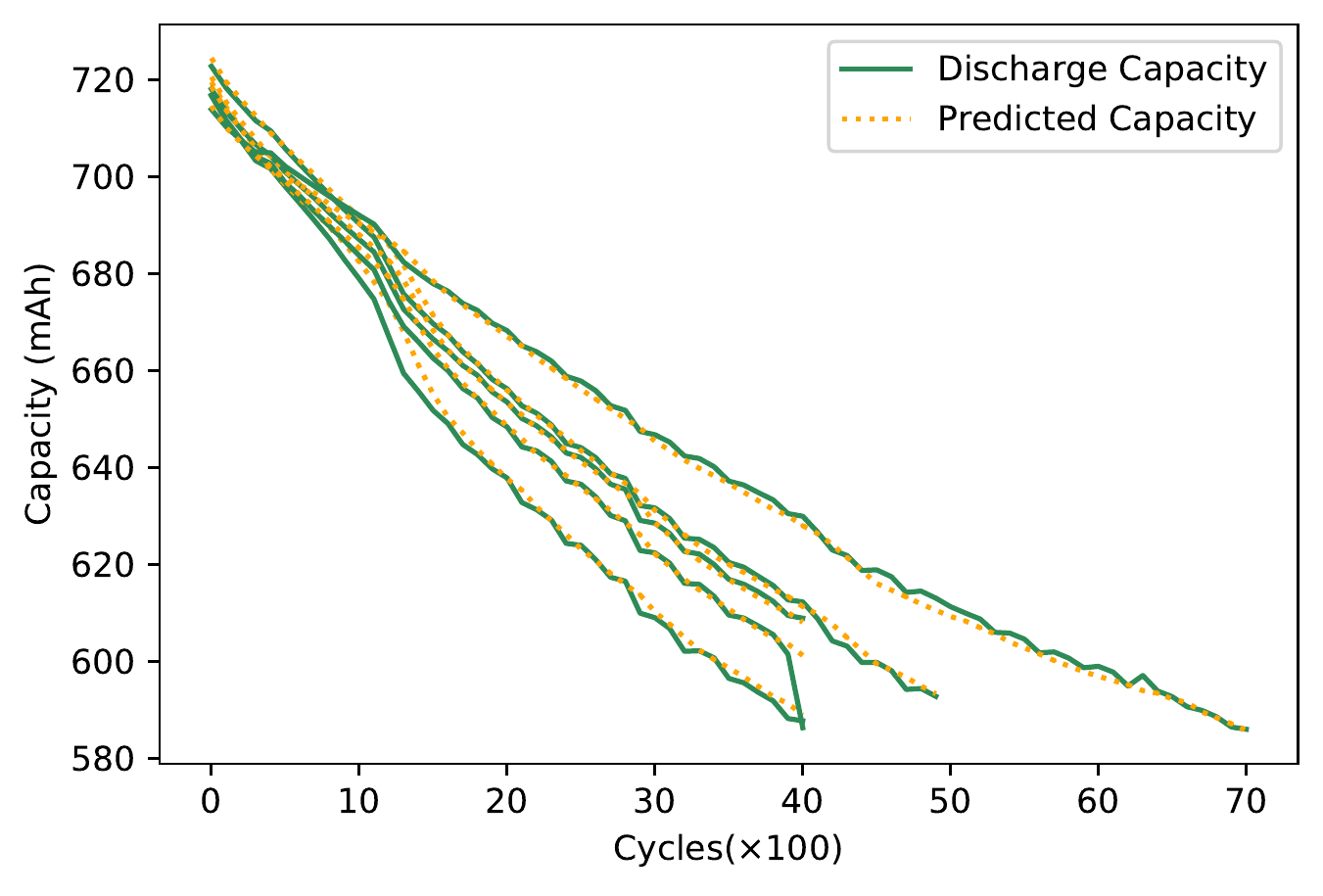}\label{fig:oxford1}}
  \subfloat[The capacity degeneration for Case 2]{\includegraphics[width=0.45\textwidth]{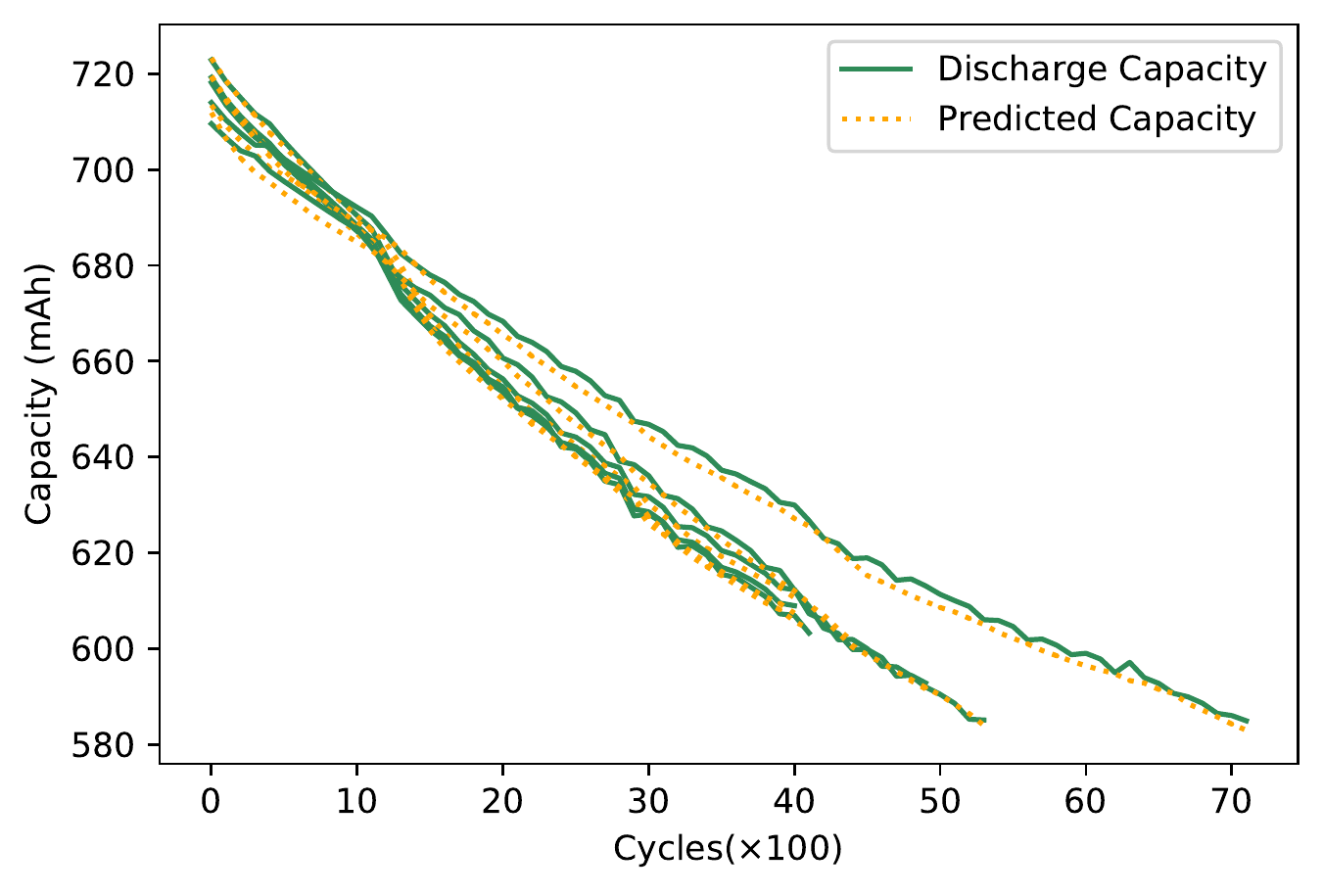}\label{fig:oxford2}}\\
  \subfloat[The capacity degeneration for Case 3]{\includegraphics[width=0.45\textwidth]{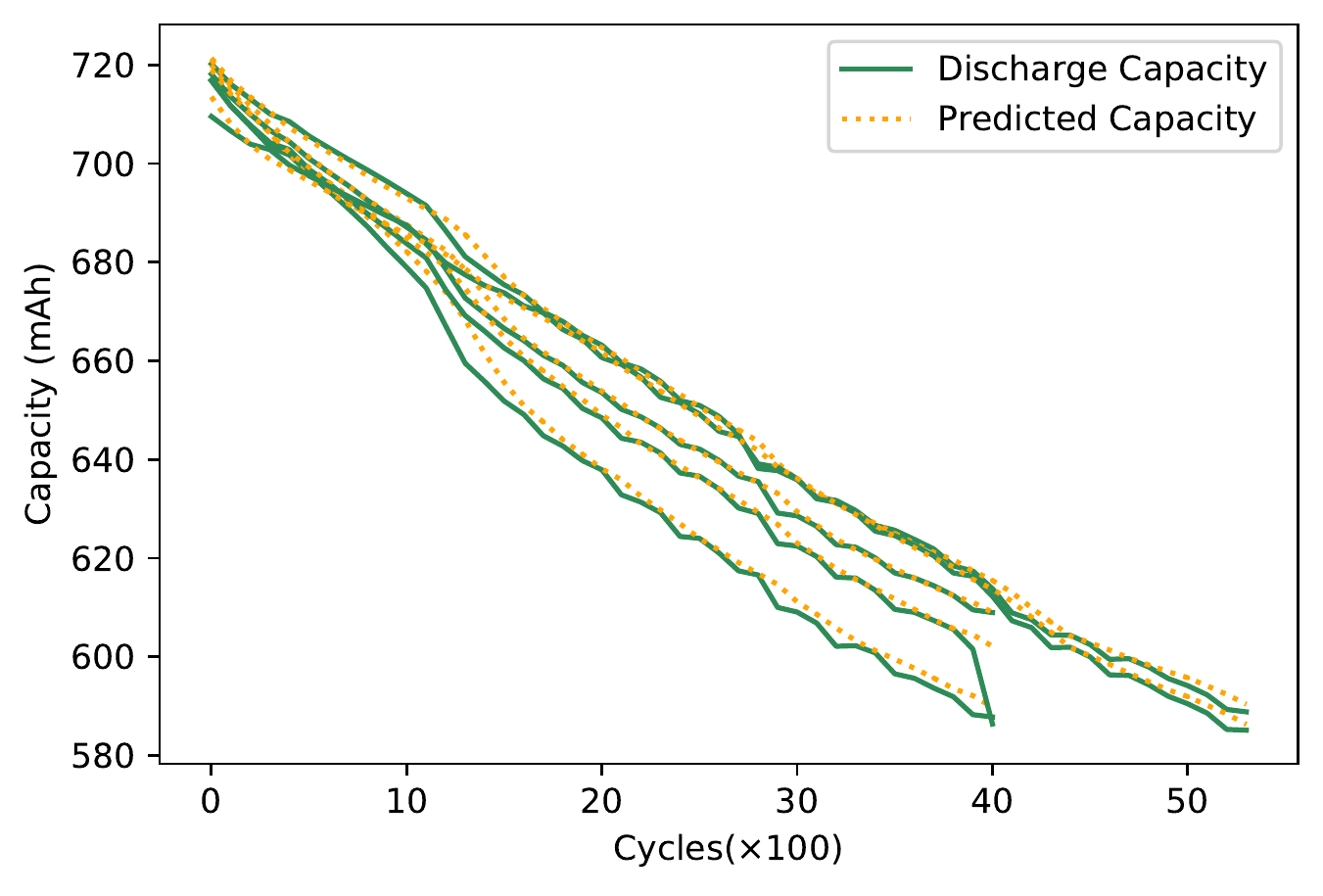}\label{fig:oxford3}}
  \subfloat[The capacity degeneration for Case 4]{\includegraphics[width=0.45\textwidth]{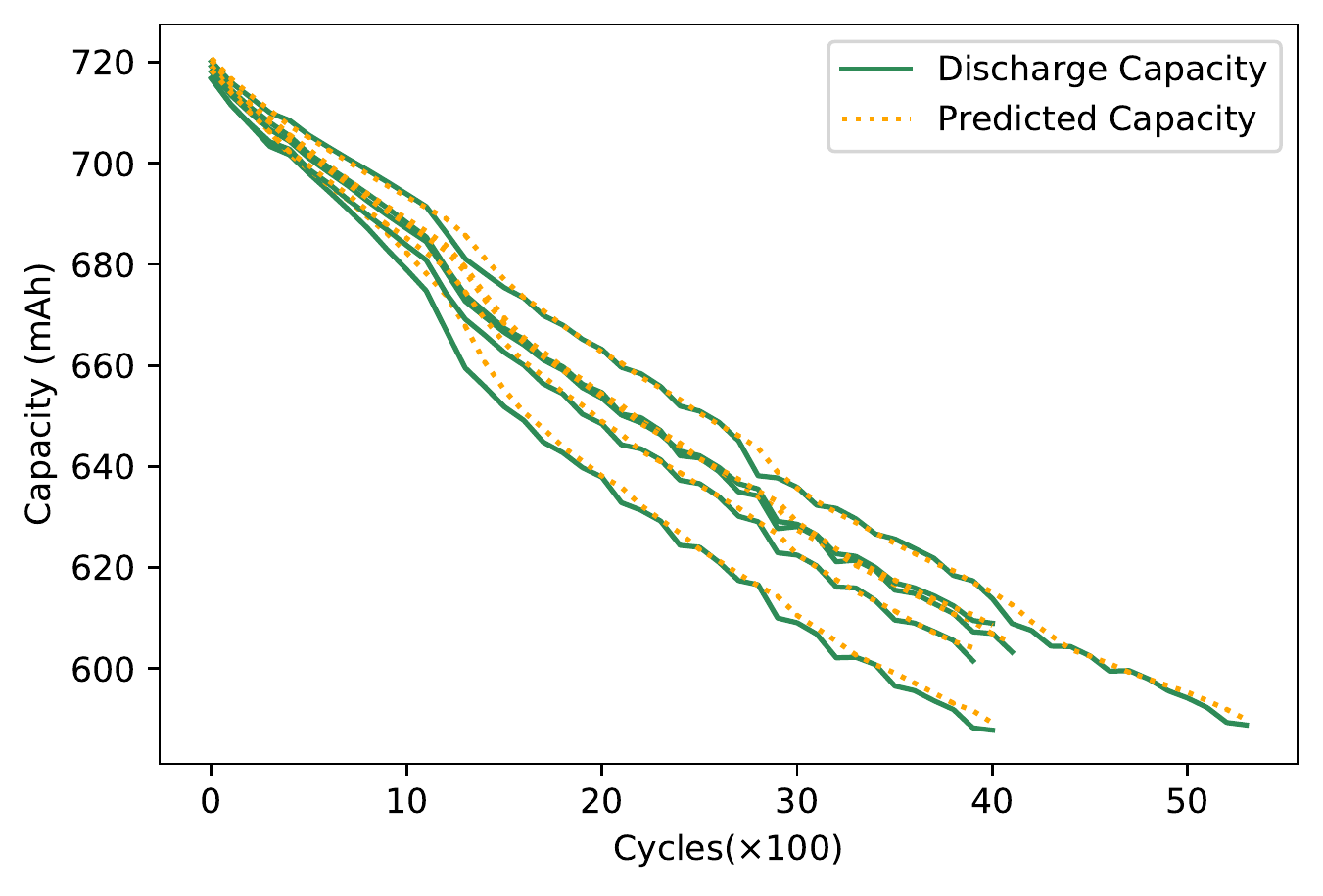}\label{fig:oxford4}}\\
  \caption{The capacity degeneration for the batteries for test in the four cases in the Oxford Battery Degradation dataset. The green lines are the measured capacity degeneration and red dots are the predicted capacity degeneration with the DDN model}\label{fig:oxford_result}
\end{figure*}

In the third performance test, the Oxford Battery Degradation Dataset\cite{birkl2017oxford,birkl2017diagnosis} is used to investigate the performance of the DDN model under dynamic load profiles. In order to benchmark with the SBC-RBFNN model\cite{she2021offline}, the SOH is predicted with the DDN model. Since SOH at each cycle is defined as the ratio between the discharge capacity and the initial discharge capacity of the first cycle of the new battery. Essentially the DDN model can be used to predict the SOH directly without modification. The streaming sensor data of the charge and discharge voltages are normalized as follows,
\begin{align}
  V_{discharge} = \frac{V_{discharge} - 2.7}{1.5}, \\
  V_{charge} = \frac{V_{charge} - 2.7}{1.5},
\end{align}
where $V_{discharge}$ is the discharge voltage data, $V_{charge}$ is the charge voltage data. The remaining settings are similar to the previous experiments. The hyper-parameters of the DDN model are listed in Table~\ref{tab:params}. 
The DDN model, in terms of RMSE, performs much better than the SBC-RBFNN model, as shown in Table~\ref{tab:result-oxford} for the four offline predictions used previously for the SBC-RBFNN model. In Case 1, No. 2, 3, 8 batteries are used for the training and No. 1, 4, 5, 6, 7 batteries for the test. In Case 2, No. 3, 4, 6 batteries are used for the training and No. 1, 2, 5, 7, 8 batteries for the test. In Case 3, No. 1, 2, 7 batteries are used for the training and No. 3, 4, 5, 6, 8 batteries for the test. In Case 4, No. 1, 7, 8 batteries are used for the training and No. 2, 3, 4, 5, 6 batteries for the test. The capacity degenerations for the four cases are shown in Fig.~\ref{fig:oxford_result} compared to the offline SBC-RBFNN model\cite{she2021offline}. The DDN model demonstrates excellent prediction accuracy in all four cases, whose average RMSE are all under 0.3$\%$.

\subsection{Attention Weights}
\begin{figure}[H]
  \centering
  \includegraphics[width=0.45\textwidth]{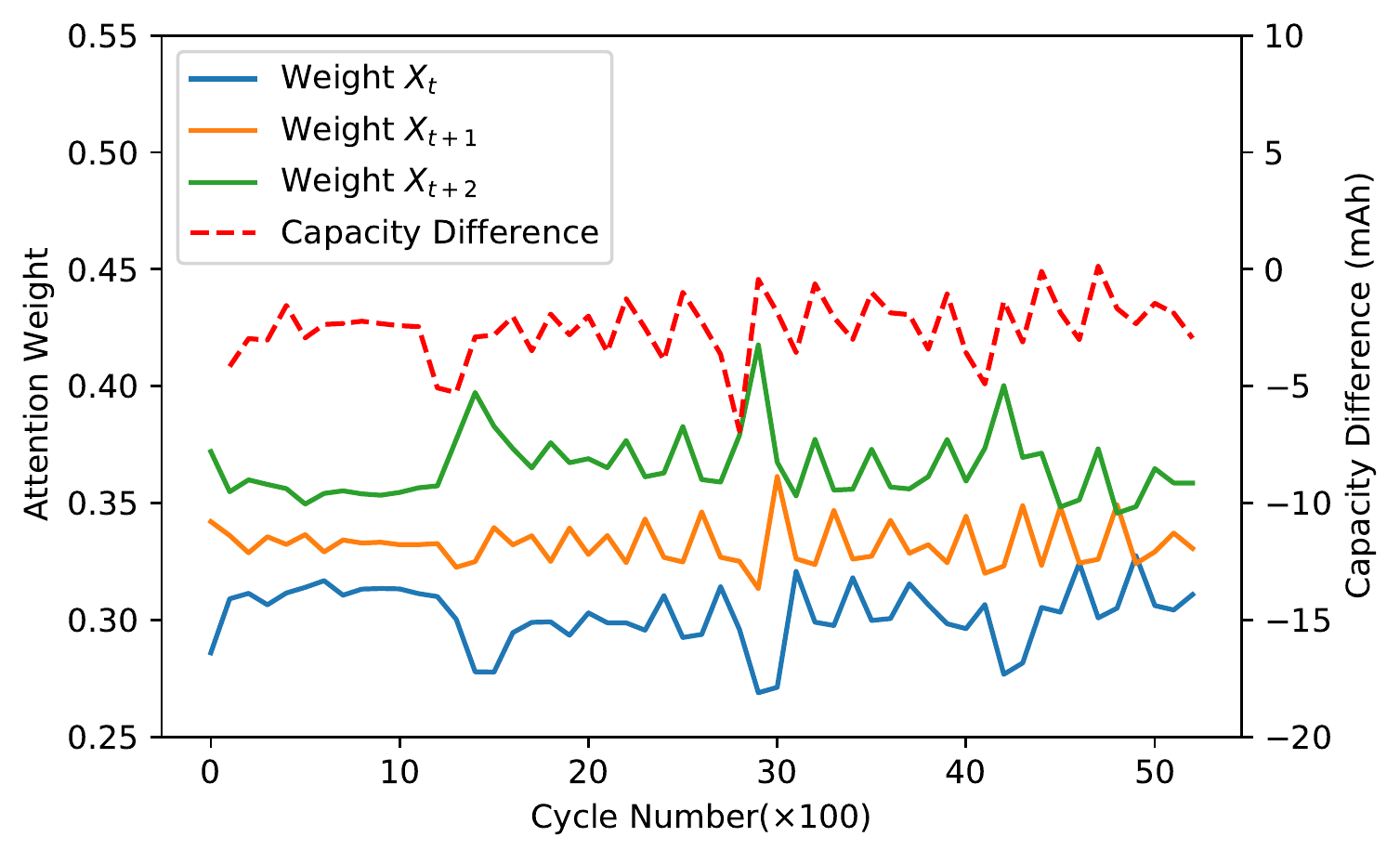}
  \caption{The curves of capacity difference and attention weight of the battery cell No.3 in Oxford Battery Degradation Dataset. The red-dashed line is the capacity difference curve of $Q_{i+1}-Q_i$. The blue, orange, and green solid lines are the $\mathbold{X}_{t}$, $\mathbold{X}_{t+1}$ and $\mathbold{X}_{t+2}$ attention weight curves in moving frames respectively.}
  \label{fig:attention_line}
\end{figure}

Moreover, Fig.~\ref{fig:attention_line} shows that the attention weights evolve with the development of capacity degeneration for the No.3 battery in the Oxford Battery Degradation Dataset.
For the prediction of $\mathbold{X}_{t+N}$, the attention weights from $\mathbold{X}_t$ to $\mathbold{X}_{t+N-1}$ in the moving frames are presented in Fig.~\ref{fig:attention_line}. For the Oxford Battery Degradation Dataset, the number of historical cycles in the moving frames, N, is 3. We can see $\mathbold{X}_{t+2}$ has the largest attention weight for the capacity degeneration prediction. Fig.~\ref{fig:attention_line} also shows the evolution of capacity difference $Q_{i+1}-Q_i$, which defines the speed of capacity degeneration. The capacity difference is strongly correlated with the attention weight, particularly the $\mathbold{X}_{t+2}$ attention weight. Since the speed of capacity degeneration is not temporally homogeneous, capacity degeneration is path-dependent\cite{sasaki2013memory, berecibar2016degradation}. The attention mechanism can catch the path-dependent memory effect in the capacity degeneration of the lithium-ion battery.

\section{Conclusion and Future Work}\label{conclusion}
This paper proposes the DDN model for the lithium-ion battery capacity forecasting using the streaming sensor data collected by the field BMS sensors. The DDN model works extremely well with the big and small datasets. The capacity degeneration prediction is very accurate. The DDN model can has comparably good performance in the small-sized NASA PCoE dataset compared to the SVR-based model and the SampEN model. For the large-sized MIT-Stanford open-access aging dataset, the DDN model has outstanding prediction accuracy for the capacity degeneration. Based on the Oxford Battery Degradation Dataset, the DDN model also have very good prediction accuracy for the capacity degeneration under the dynamic load profiles. The DDN model has the following advantages,
\begin{enumerate}
  \item The network uses streaming sensor data directly without heavy data preprocessing. The DDN model is fully data-driven and extracts the temporal degeneration pattern in the streaming sensor data. 
  \item The DDN model has very high efficiency and is adaptive to the streaming sensor data. The model training and its hyper-parameters optimization are very fast.
  \item The DDN model has very high performance and can easily be implemented in the IoT framework.
  \item The DDN model is highly accurate and robust.
\end{enumerate}
The streaming sensor data extensively exist in the medical, mechanical, transportation, and power systems. In the future, we will further explore the applications of the model in battery management systems such as fast-charging, health monitoring, and fault detection of batteries, {\it etc}. Currently, the devices in the smart grid all have good processing capability. Moreover, given the advantages and efficiency, the DDN model has promising applications in edge computing for the cloud battery management system.

\bibliographystyle{unsrt}
\bibliography{references}  

\end{document}